\documentclass[aps,prl,preprintnumbers,amsmath,amssymb,superscriptaddress]{revtex4}%

\usepackage{graphicx}
\usepackage{dcolumn}
\usepackage{bm}
\usepackage{color}

\begin{document}

\title{Evidence for granularity, anisotropy and lattice distortions in cuprate
superconductors and their implications}
\author{H.~Keller}
\affiliation{Physik-Institut der Universit\"{a}t Z\"{u}rich,
Winterthurerstrasse 190, CH-8057, Switzerland}
\author{T.~Schneider}
\affiliation{Physik-Institut der Universit\"{a}t Z\"{u}rich,
Winterthurerstrasse 190, CH-8057, Switzerland}
%

\begin{abstract}
Granularity, anisotropy, local lattice distortions and their
dependence on dopant concentration appear to be present in all
cuprate superconductors, interwoven with the microscopic
mechanisms responsible for superconductivity. Here we review
anisotropy and penetration depth measurements to reassess the
evidence for granularity, as revealed by the notorious rounded
phase transition, the evidence for the three dimensional nature of
superconductivity, uncovered by the doping dependence of
transition temperature and anisotropy, and to reassess the
relevance of the electron-lattice coupling, emerging from the
oxygen isotope effects.
\end{abstract}
\maketitle

\bigskip
\begin{center}
To appear in the proceedings of \ Symmetry and Heterogeneity in
High Temperature Superconductors, Erice-Sicily: 4-10 October 2003
\end{center}

\begin{figure}[tbp]
\centering
\includegraphics[totalheight=6cm]{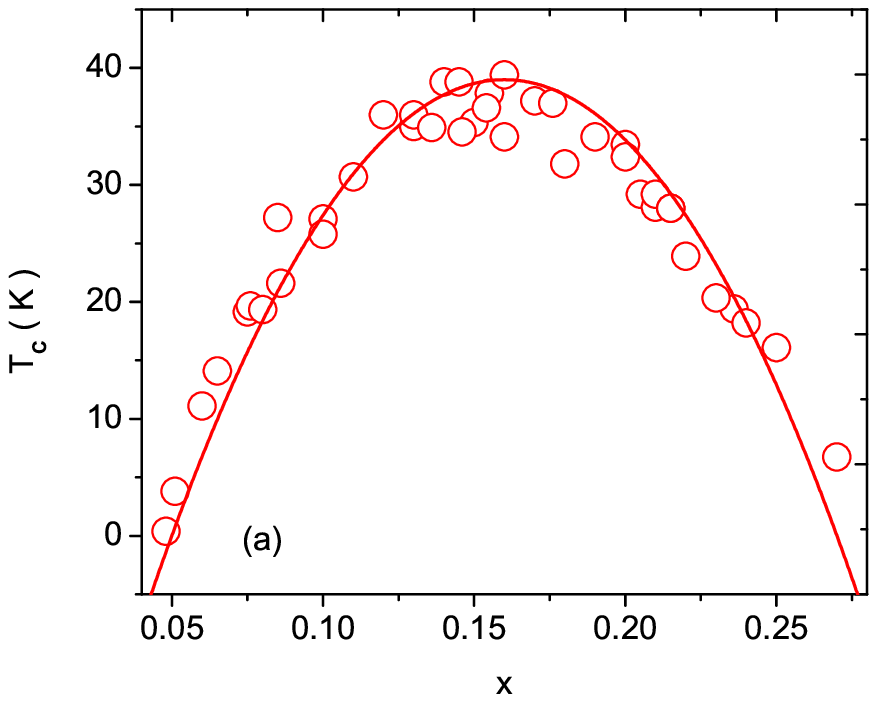}
\includegraphics[totalheight=6cm]{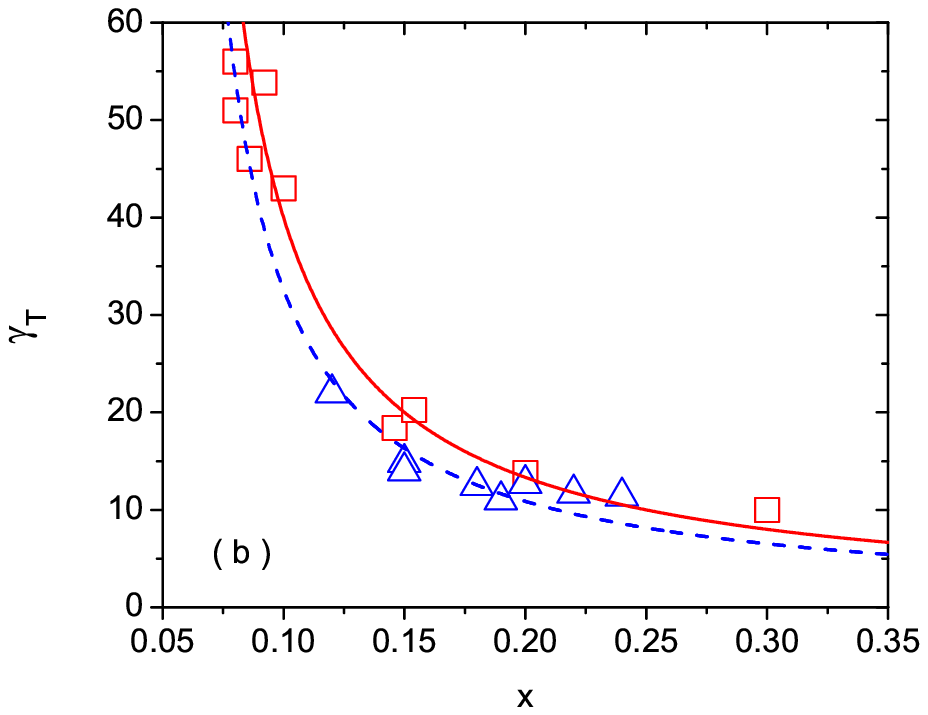}
\caption{(a) Variation of $T_{c}$ for La$_{2-x}$Sr$_{x}$CuO$_{4}$.
Experimental data taken from \protect\cite
{suzuki,nakamura,fukuzumi,willemin,kimura,sasagawa,hoferdis,shibauchi,panagopoulos}.
The solid line is Eq.(\ref{eq1}) with $T_{c}\left( x_{m}\right)
=39$K. (b) $\gamma _{T}$ versus $x$ for
La$_{2-x}$Sr$_{x}$CuO$_{4}$. The squares are the experimental data
for $\gamma _{T_{c}}$ \protect\cite
{suzuki,nakamura,willemin,sasagawa,hoferdis} and the triangles for
$\gamma _{T=0}$ \protect\cite{shibauchi,panagopoulos}. The solid
curve and dashed lines are Eq.(\ref{eq2}) with  $\gamma
_{T_{c},0}=2$ and $\gamma _{T=0,0}=1.63$.} \label{fig1}
\end{figure}

Establishing and understanding the phase diagram of cuprate
superconductors in the temperature - dopant concentration plane is
one of the major challenges in condensed matter physics.
Superconductivity is derived from the insulating and
antiferromagnetic parent compounds by partial substitution of ions
or by adding or removing oxygen. For instance La$_{2}$CuO$_{4}$
can be doped either by alkaline earth ions or oxygen to exhibit
superconductivity. The empirical phase diagram of
La$_{2-x}$Sr$_{x}$CuO$_{4}$ \cite
{suzuki,nakamura,fukuzumi,willemin,kimura,sasagawa,hoferdis,shibauchi,panagopoulos}
depicted in Fig.\ref{fig1}a shows that after passing the so called
underdoped limit $\left( x_{u}\approx 0.05\right) $, $T_{c}$
reaches its maximum value $T_{c}\left( x_{m}\right) $ at
$x_{m}\approx 0.16$. With further increase of $x $, $T_{c}$
decreases and finally vanishes in the overdoped limit
$x_{o}\approx 0.27$. This phase transition line is thought to be a
generic property of cuprate superconductors \cite{tallon} and is
well described by the empirical relation
\begin{equation}
T_{c}\left( x\right) =T_{c}\left( x_{m}\right) \left( 1-2\left(
\frac{x}{x_{m}}-1\right) ^{2}\right) =\frac{2T_{c}\left(
x_{m}\right) }{x_{m}^{2}}\left( x-x_{u}\right) \left(
x_{o}-x\right) ,  \label{eq1}
\end{equation}
proposed by Presland \emph{et al}.\cite{presland}. Approaching the
endpoints along the $x$-axis, La$_{2-x}$Sr$_{x}$CuO$_{4}$
undergoes at zero temperature doping tuned quantum phase
transitions. As their nature is concerned, resistivity
measurements reveal a quantum superconductor to insulator (QSI)
transition in the underdoped limit\cite
{momono,polen,book,klosters,tshk,tsphysB,parks} and in the
overdoped limit a quantum superconductor to normal state (QSN)
transition\cite{momono}.

Another essential experimental fact is the doping dependence of
the anisotropy. In tetragonal cuprates it is defined as the ratio
$\gamma =\xi _{ab}/\xi _{c}$ of the correlation lengths parallel
$\left( \xi _{ab}\right) $ and perpendicular $\left( \xi
_{c}\right) $ to CuO$_{2}$ layers ($ab$-planes). In the
superconducting state it can also be expressed as the ratio
$\gamma =\lambda _{c}/\lambda _{ab}$ of the London penetration
depths due to supercurrents flowing perpendicular ($\lambda _{c}$
) and parallel ($\lambda _{ab}$ ) to the $ab$-planes. Approaching
a nonsuperconductor to superconductor transition $\xi $ diverges,
while in a superconductor to nonsuperconductor transition $\lambda
$ tends to infinity. In both cases, however, $\gamma $ remains
finite as long as the system exhibits anisotropic but genuine 3D
behavior. There are two limiting cases: $\gamma =1$ characterizes
isotropic 3D- and $\gamma =\infty $ 2D-critical behavior. An
instructive model where $\gamma $ can be varied continuously is
the anisotropic 2D Ising model\cite{onsager}. When the coupling in
the $y$ direction goes to zero, $\gamma =\xi _{x}/\xi _{y}$
becomes infinite, the model reduces to the 1D case, and $T_{c}$
vanishes. In the Ginzburg-Landau description of layered
superconductors the anisotropy is related to the interlayer
coupling. The weaker this coupling is, the larger $\gamma $ is.
The limit $\gamma =\infty $ is attained when the bulk
superconductor corresponds to a stack of independent slabs of
thickness $d_{s}$. With respect to experimental work, a
considerable amount of data is available on the chemical
composition dependence of $\gamma $. At $T_{c}$ it can be inferred
from resistivity ($\gamma =\xi _{ab}/\xi _{c}=\sqrt{\rho
_{ab}/\rho _{c}}$) and magnetic torque measurements, while in the
superconducting state it follows from magnetic torque and
penetration depth ($\gamma =\lambda _{c}/\lambda _{ab}$) data. In
Fig. \ref{fig1}b we displayed the doping dependence of $1/\gamma
_{T}$ evaluated at $T_{c}$ ($\gamma _{T_{c}}$) and $T=0$ ($\gamma
_{T=0}$). As the dopant concentration is reduced, $\gamma
_{T_{c}}$ and $\gamma _{T=0}$ increase systematically, and tend to
diverge in the underdoped limit. Thus the temperature range where
superconductivity occurs shrinks in the underdoped regime with
increasing anisotropy. This competition between anisotropy and
superconductivity raises serious doubts whether 2D mechanisms and
models, corresponding to the limit $\gamma _{T}=\infty $, can
explain the essential observations of superconductivity in the
cuprates. From Fig. \ref{fig1}b it is also seen that $\gamma
_{T}\left( x\right) $ is well described by
\begin{equation}
\gamma _{T}\left( x\right) =\frac{\gamma _{T,0}}{x-x_{u}},
\label{eq2}
\end{equation}
where $\gamma _{T,0}$ is the quantum critical amplitude. Having
also other cuprate families in mind, it is convenient to express
the dopant concentration in terms of $T_{c}$. From Eqs.(\ref{eq1})
and(\ref{eq2}) we obtain the correlation between $T_{c}$ and
$\gamma _{T}$:
\begin{equation}
\frac{T_{c}\left( x\right) }{T_{c}\left( x_{m}\right) }=1-\left(
\frac{\gamma _{T}\left( x_{m}\right) }{\gamma _{T}\left( x\right)
}-1\right) ^{2},\ \ \gamma _{T}\left( x_{m}\right) =\frac{\gamma
_{T,0}}{x_{m}-x_{u}} \label{eq3}
\end{equation}

\begin{figure}[tbp]
\centering
\includegraphics[totalheight=6cm]{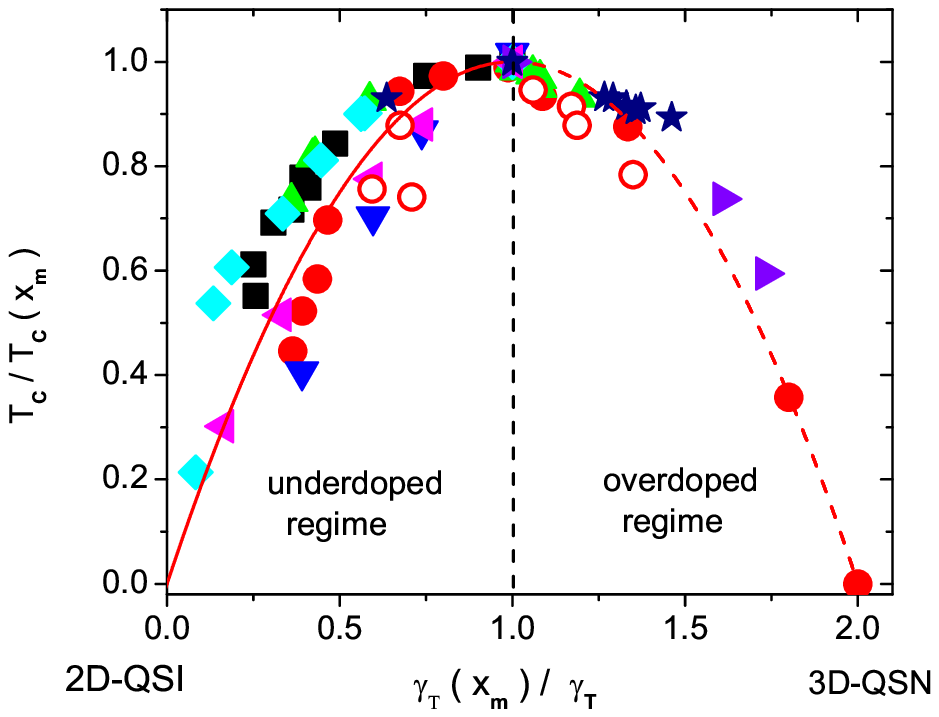}
\caption{$T_{c}\left( x\right) /T_{c}\left( x_{m}\right) $ versus
$\gamma _{T}\left( x_{m}\right) /$ $\gamma _{T}\left( x\right) $
for various cuprate families: La$_{2-x}$Sr$_{x}$CuO$_{4}$
($\bullet $, $T_{c}\left( x_{m}\right) =37$K, $\gamma
_{T_{c}}\left( x_{m}\right) =20$) \protect\cite
{suzuki,nakamura,willemin,sasagawa,hoferdis}\ , ($\bigcirc $,
$T_{c}\left( x_{m}\right) =37$K, $\gamma _{T=0}\left( x_{m}\right)
=14.9$) \protect\cite {shibauchi,panagopoulos},
HgBa$_{2}$CuO$_{4+\delta }$ ($\blacktriangle $, $T_{c}\left(
x_{m}\right) =95.6$K, $\gamma _{T_{c}}\left( x_{m}\right) =27$)
\protect\cite{hoferhg}, Bi$_{2}$Sr$_{2}$CaCu$_{2}$O$_{8+\delta }$
($\bigstar $, $T_{c}\left( x_{m}\right) =84.2$K, $\gamma
_{T_{c}}\left( x_{m}\right) =133$) \protect\cite{watauchi},
YBa$_{2}$Cu$_{3}$O$_{7-\delta }$ ($\blacklozenge $, $T_{c}\left(
x_{m}\right) =92.9$K, $\gamma _{T_{c}}\left( x_{m}\right) =8$)
\protect\cite{chien123},
YBa$_{2}$(Cu$_{1-y}$Fe$_{y}$)$_{3}$O$_{7-\delta }$ ($\blacksquare
$, $T_{c}\left( x_{m}\right) =92.5$K, $\gamma _{T_{c}}\left(
x_{m}\right) =9$)\protect\cite{chienfe},
Y$_{1-y}$Pr$_{y}$Ba$_{2}$Cu$_{3}$O$_{7-\delta }$
($\blacktriangledown $, $T_{c}\left( x_{m}\right) =91$K, $\gamma
_{T_{c}}\left( x_{m}\right) =9.3$)\protect\cite{chienpr},
BiSr$_{2}$Ca$_{1-y}$Pr$_{y}$Cu$_{2}$O$_{8}$ ($\blacktriangleleft
$, $T_{c}\left( x_{m}\right) =85.4$K, $\gamma _{T=0}\left(
x_{m}\right) =94.3$)\protect\cite{sun} and YBa$_{2}$(Cu$_{1-y}$
Zn$_{y}$)$_{3}$O$_{7-\delta }$ ($\blacktriangleright $,
$T_{c}\left( x_{m}\right) =92.5$K, $\gamma _{T=0}\left(
x_{m}\right) =9$)\protect\cite {panagopzn}. The solid and dashed
curves are Eq.(\ref{eq3}), marking the flow from the maximum
$T_{c}$ to QSI and QSN criticality, respectively.} \label{fig2}
\end{figure}
Provided that this empirical correlation is not merely an artefact
of La$_{2-x}$Sr$_{x}$CuO$_{4}$, it gives a universal perspective
on the interplay of anisotropy and superconductivity, among the
families of cuprates, characterized by $T_{c}\left( x_{m}\right) $
and $\gamma _{T}\left( x_{m}\right) $. For this reason it is
essential to explore its generic validity. In practice, however,
there are only a few additional compounds, including
HgBa$_{2}$CuO$_{4+\delta }$\cite{hoferhg}, for which the dopant
concentration can be varied continuously throughout the entire
doping range. It is well established, however, that the
substitution of magnetic and nonmagnetic impurities depresses
$T_{c}$ of cuprate superconductors very
effectively\cite{xiao,tarascon}. To compare the doping and
substitution driven variations of the anisotropy, we depicted in
Fig. \ref{fig2} the plot $T_{c}\left( x\right) /T_{c}\left(
x_{m}\right) $ versus $\gamma _{T}\left( x_{m}\right) /$ $\gamma
_{T}\left( x\right) $ for a variety of cuprate families. The
collapse of the data on the parabola, which is the empirical
relation (\ref{eq3}), reveals that this scaling form appears to be
universal. Thus, given a family of cuprate superconductors,
characterized by $T_{c}\left( x_{m}\right) $ and $\gamma
_{T}\left( x_{m}\right) $, it gives a universal perspective on the
interplay between anisotropy and superconductivity.

\begin{figure}[tbp]
\centering
\includegraphics[totalheight=6cm]{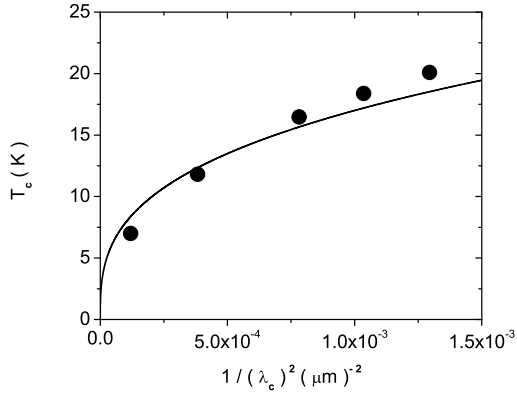}
\caption{$T_{c}$ versus $1/\lambda _{c}^{2}\left( T=0\right) $ for
heavily underdoped YBa$_{2}$Cu$_{3}$O$_{7-\delta }$ single
crystals, taken from Hosseini\protect\cite{hosseini}. The solid
line is $T_{c}$ $=170\left( 1/\lambda _{c}^{2}\left( T=0\right)
\right) ^{1/3}$ and indicates the consistency with the 2D-QSI
scaling relation $T_{c}\propto \left( 1/\lambda _{c}^{2}\left(
0\right) \right) ^{z/(z+2)}$ and $z=1$.} \label{fig3}
\end{figure}

Close to 2D-QSI criticality various properties are not independent
but related by\cite{polen,book,klosters,tshk,tsphysB,parks}
\begin{equation}
T_{c}=\frac{\Phi _{0}^{2}R_{2}}{16\pi
^{3}k_{B}}\frac{d_{s}}{\lambda _{ab}^{2}\left( 0\right) }\propto
\gamma _{T}^{-z}\propto \delta ^{z\overline{\nu }},  \label{eq4}
\end{equation}
where $k_{B}$ is the Boltzmann constant, and $\Phi _{0}$ the
elementary flux quantum. $\lambda _{ab}\left( 0\right) $ is the
zero temperature in-plane penetration depth, $z$ is the dynamic
critical exponent, $d_{s}$ the thickness of the sheets, and
$\overline{\nu }$ the correlation length critical exponent of the
2D-QSI transitions. $\delta $ measures the distance from the
critical point along the $x$ axis (see Fig.\ref{fig1}a, and
$R_{2}$ is a universal number. Since $T_{c}\propto d_{s}/\lambda
_{ab}^{2}\left( 0\right) \propto n_{s}^{\Box }$, where
$n_{s}^{\Box }$ is the aerial superfluid density, is a
characteristic 2D property, it also applies to the onset of
superfluidity in $^{4}$He films adsorbed on disordered substrates,
where it is well confirmed\cite{crowell}. A great deal of
experimental work has also been done in cuprates on the so called
Uemura plot, revealing an empirical correlation between $T_{c}$
and $d_{s}/\lambda _{ab}^{2}\left( 0\right) $\cite{uemura}.
Approaching 2D-QSI criticality, the data of a given family tends
to fall on a straight line, consistent with Eq.(\ref{eq4}).
Differences in the slope reflect the family dependent value of
$d_{s}$, the thickness of the sheets, becoming independent in the
2D limit\cite {book,klosters,tshk,tsphysB,parks}. The relevance of
$d_{s}$ was also confirmed in terms of the relationship between
the isotope effect on $T_{c}$ and $1/\lambda
_{ab}^{2}$\cite{tshk,tsiso}. Moreover, together with the scaling
form (\ref{eq4}) the empirical relation (\ref{eq1}) implies 2D-QSI
and 3D-QSN transitions with $z=1$, while the empirical relation
for the anisotropy (Eqs.(\ref{eq2}) and (\ref{eq3})), require
$\overline{\nu }=1$ at the 2D-QSI critical point. Thus, the
empirical correlations point to a 2D-QSI transition with $z=1$ and
$\overline{\nu }=1$.These estimates coincide with the theoretical
prediction for a 2D disordered bosonic system with long-range
Coulomb interactions, where $z=1$ and $\overline{\nu }\simeq
1$\cite{mpafisher,ca,herbutz1}. Here the loss of superfluidity is
due to the localization of the pairs, which ultimately drives the
transition. From the scaling relation (\ref{eq4}) it is seen that
measurements of the out of plane penetration depth of sufficiently
underdoped systems allow to estimate the dynamic critical exponent
$z$ directly, in terms of $T_{c}\propto \left( 1/\lambda
_{c}^{2}\left( 0\right) \right) ^{z/\left( z+2\right) }$, which
follows from Eq.(\ref{eq4}) with $\gamma _{T}=\lambda _{c}\left(
0\right) /\lambda _{ab}\left( 0\right) $. In Fig.\ref{fig3} we
displayed the data of Hosseini\cite{hosseini} for heavily
underdoped YBa$_{2}$Cu$_{3}$O$_{7-\delta }$ single crystals. The
solid line is $T_{c}$ $=170\left( 1/\lambda _{c}^{2}\left(
T=0\right) \right) ^{1/3}$ and uncovers the consistency with the
2D-QSI scaling relation $T_{c}\propto \left( 1/\lambda
_{c}^{2}\left( 0\right) \right) ^{z/\left( z+2\right) }$ with
$z=1$.

We have seen that the doping tuned flow to the 2D-QSI critical
point is associated with a depression of $T_{c}$ and an
enhancement of $\gamma _{T}$. It implies that whenever a QSI
transition is approached, a no vanishing $T_{c}$ is inevitably
associated with an anisotropic but 3D condensation mechanism,
because $\gamma _{T}$ is finite for $T_{c}>0$ (see
Figs.\ref{fig1}b and \ref{fig2}). This represents a serious
problem for 2D models\cite {anderson} as candidates to explain
superconductivity in the cuprates, and serves as a constraint on
future work toward a complete understanding. Note that the vast
majority of theoretical models focus on a single Cu-O plane, i.e.,
on the limit of zero intracell and intercell $c$-axis coupling.

\begin{figure}[tbp]
\centering
\includegraphics[totalheight=6cm]{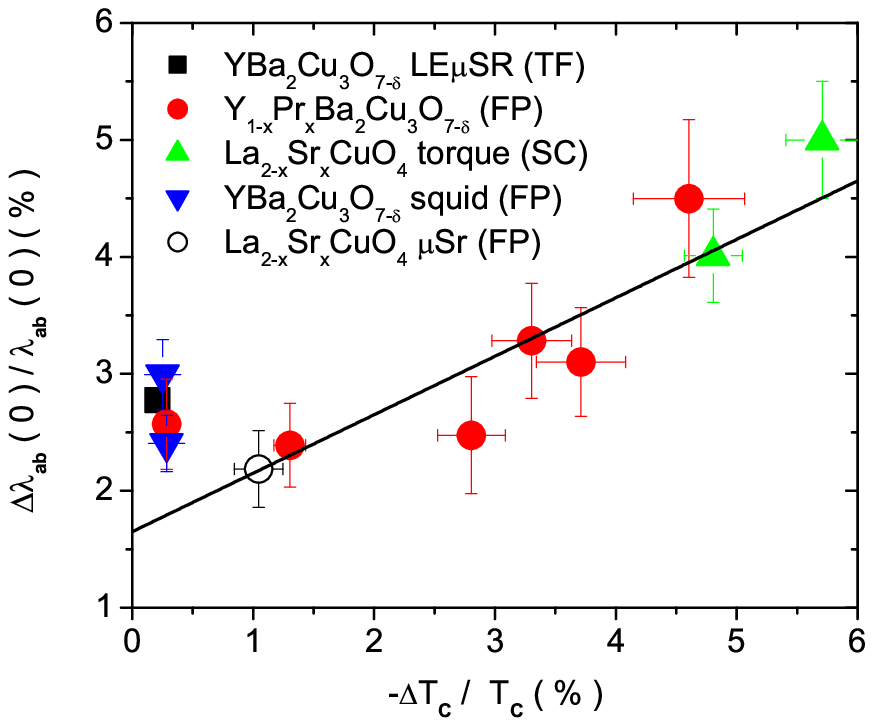}
\caption{Data for the oxygen isotope effect in underdoped
La$_{2-x}$Sr$_{x}$CuO$_{4}$($\bigcirc $:
x=0.15\protect\cite{zhao1}, $\blacktriangle $:x=0.08, 0.086
\protect\cite {hofer214},
Y$_{1-x}$Pr$_{x}$Ba$_{2}$Cu$_{3}$O$_{7-\delta }$ ($\bullet $: x=0,
0.2, 0.3, 0.4)\protect\cite{zhao1,khasanov123pr,rksite} and
YBa$_{2}$Cu$_{3}$O$_{7-\delta }$ ($\blacktriangledown $
\protect\cite{zhao1}, $\blacksquare $\protect\cite {khasanov123f})
\ in terms of $\Delta \left( \lambda _{ab}\left( 0\right) \right)
/$ $\lambda _{ab}\left( 0\right) $ versus -$\Delta T_{c}/T_{c}$.
The solid line indicates the flow to 2D-QSI criticality and
provides with Eq.(\ref{eq5}) an estimate for the oxygen isotope
effect on $d_{s}$, namely $\Delta d_{s}/d_{s}=3.3(4)\%$.}
\label{fig4}
\end{figure}

\begin{figure}[tbp]
\centering
\includegraphics[totalheight=6cm]{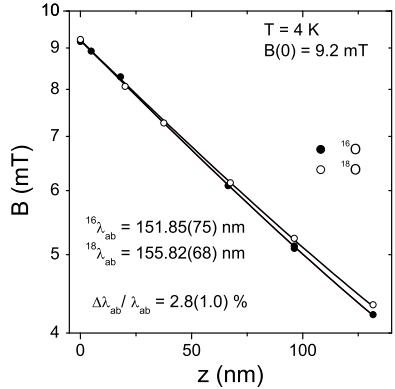}
\caption{Magnetic field penetration profiles $B(z)$ for a $^{16}$O
substituted (closed symbols) and a $^{18}$O substituted (open
symbols) YBa$_{2}$Cu$_{3}$O$_{7-\delta }$-film measured in the
Meissner state at 4 K and an external field of 9.2 mT, applied
parallel to the surface of the film. The data are shown for
implantation energies 3, 6, 10, 16, 22, and 29 keV starting from
the surface of the sample. Solid curves are best fits to
$B(z)=B_{0}\cosh[(t-z)/\lambda _{ab}]\cosh(t/\lambda _{ab})$. This
is the form of the classical exponential field decay in the
Meissner state $B(z)=B_{0}\exp(-z/\lambda _{ab})$, modified for a
film with thickness $2t$ with flux penetrating from both sides.
Taken from Khasanov \emph{et al.}\protect\cite{khasanov123f} }
\label{fig5}
\end{figure}

Since Eq.(\ref{eq4}) is universal, it also implies that the
changes $\Delta T_{c}$, $\Delta d_{s}$ and $\Delta \left(
1/\lambda _{ab}^{2}\left( T=0\right) \right) $, induced by
pressure or isotope exchange are not independent, but related by
\begin{equation}
\frac{\Delta T_{c}}{T_{c}}=\frac{\Delta d_{s}}{d_{s}}+\frac{\Delta
\left( 1/\lambda _{ab}^{2}\left( 0\right) \right) }{\left(
1/\lambda _{ab}^{2}\left( 0\right) \right) }=\frac{\Delta
d_{s}}{d_{s}}-2\frac{\Delta \left( \lambda _{ab}\left( 0\right)
\right) }{\lambda _{ab}\left( 0\right) }. \label{eq5}
\end{equation}
In particular, for the oxygen isotope effect ($^{16}$O vs.
$^{18}$O) of a physical quantity $X$ \ the relative isotope shift
is defined as $\Delta X/X=(^{18}X-^{16}X)/^{18}X$. In
Fig.\ref{fig4} we show the data for the oxygen isotope effect in
La$_{2-x}$Sr$_{x}$CuO$_{4}$\cite{hofer214,zhao1},
Y$_{1-x}$Pr$_{x}$Ba$_{2}$Cu$_{3}$O$_{7-\delta }$\cite
{zhao1,khasanov123pr,rksite} and YBa$_{2}$Cu$_{3}$O$_{7-\delta
}$\cite {zhao1,khasanov123f}, extending from the underdoped to the
optimally doped regime, in terms of $\Delta \left( \lambda
_{ab}\left( 0\right) \right) /$ $\lambda _{ab}\left( 0\right) $
versus $\Delta T_{c}/T_{c}$. It is evident that there is a
correlation between the isotope effect on $T_{c}$ and $\lambda
_{ab}\left( 0\right) $ which appears to be universal for all
cuprate families. Indeed, the solid line indicates the flow to the
2D-QSI transition and provides with Eq.(\ref{eq5}) an estimate for
the oxygen isotope effect on $d_{s}$, namely $\Delta
d_{s}/d_{s}=3.3(4)\%$. Approaching optimum doping, this
contribution renders the isotope effect on $T_{c}$ considerably
smaller than that on $\lambda _{ab}\left( 0\right) $. As shown in
Fig.\ref {fig5}, even in nearly optimally doped
YBa$_{2}$Cu$_{3}$O$_{7-\delta }$, where $\Delta
T_{c}/T_{c}=-0.26(5)\%$, a substantial isotope effect on the
in-plane penetration depth, $\Delta \lambda _{ab}\left( 0\right)
/\lambda _{ab}\left( 0\right) =-2.8(1.0)\%$, has been established
by direct observation, using the novel low-energy muon-spin
rotation technique\cite {khasanov123f}. Note that these findings
have been obtained using various experimental techniques on
powders, thin films and single crystals.

Since upon oxygen isotope exchange the lattice parameters remain
essentially unaffected\cite{raffa,conder}, the substantial isotope
effect on the in-plane penetration depth uncovers the coupling
between local lattice distortions and superfluidity and the
failure of the Migdal-Eliashberg (ME) treatment of the
electron-phonon interaction, predicting, $1/\lambda ^{2}\left(
0\right) $, to be independent of the ionic masses\cite{migdal}.
Evidence for this coupling emerges from the oxygen isotope effect
on $d_{s}$, the thickness of the superconducting sheets, upon
isotope exchange, while the lattice parameters remain unaffected.
Indeed, the relative shift, $\Delta d_{s}/d_{s}\approx 3.3(4)\%$,
apparent in Fig.\ref{fig3}, implies local distortions of oxygen
degrees of freedom, which do not modify the lattice parameters,
and are coupled to the superfluid.

Further evidence for this coupling emerges from the isotope effect
on the granularity of the cuprates\cite{varenna,tsrkhk}. Recently,
it has been shown that the notorious rounding of the
superconductor to normal state transition is fully consistent with
a finite size effect, revealing that bulk cuprate superconductors
break into nearly homogeneous superconducting grains of rather
unique extent\cite{book,varenna,tsrkhk,bled,tsdc}. Even evidence
for their surface and edge contributions to specific heat and
penetration depth has been established\cite{tserice}. A
characteristic feature of a finite size effect in the temperature
dependence of the in-plane penetration depth $\lambda _{ab}$ is
the occurrence of an inflection point giving rise to an extremum
in $d\left( \lambda _{ab}^{2}\left( T=0\right) /\lambda
_{ab}^{2}\left( T\right) \right) /dT$ at $T_{p}$. Here $\lambda
_{ab}^{2}\left( T_{p}\right) $, $T_{p}$ and the length $L_{c}$ of
the grains along the c-axis are related by\cite
{varenna,tsrkhk,bled,tsdc,tserice}
\begin{equation}
\frac{1}{\lambda _{ab}^{2}\left( T_{p}\right) }=\frac{16\pi
^{3}k_{B}T_{p}}{\Phi _{0}^{2}L_{c}}.  \label{eq7}
\end{equation}
Recently we explored the effect of oxygen isotope exchange in
Y$_{1-x}$Pr$_{x}$Ba$_{2}$Cu$_{3}$O$_{7-\delta }$ on $L_{c}$ by
means of in-plane penetration depth measurements\cite{tsrkhk}.
Note that the shifts are not independent but according to
Eq.(\ref{eq7}) related by
\begin{equation}
\frac{\Delta L_{c}}{L_{c}}=\frac{\Delta
T_{p_{c}}}{T_{p_{c}}}+\frac{\Delta \lambda _{ab}^{2}\left(
T_{p_{c}}\right) }{\lambda _{ab}^{2}\left( T_{p_{c}}\right) }.
\label{eq8}
\end{equation}
From the resulting estimates, listed in Table I, several
observations emerge. First, $L_{c}$ increases systematically with
reduced $T_{p_{c}}$. Second, $L_{c}$ grows with increasing $x$ and
upon isotope exchange ($^{16}$O, $^{18}$O). Third, the relative
shift of $T_{p_{c}}$ is very small. This reflects the fact that
the change of $L_{c}$ is essentially due to the superfluid, probed
in terms of $\lambda _{ab}^{2}$. Accordingly, $\Delta
L_{c}/L_{c}\approx \Delta \lambda _{ab}^{2}/\lambda _{ab}^{2}$ for
$x=0,\ 0.2 $ and $0.3$.

\bigskip

\begin{center}
\begin{tabular}{|c|c|c|c|}
\hline x & 0 & 0.2 & 0.3 \\ \hline $\Delta L_{c}/L_{c}$ & 0.12(5)
& 0.13(6) & 0.16(5) \\ \hline $\Delta T_{p_{c}}/T_{p_{c}}$ &
-0.000(2) & -0.015(3) & -0.021(5) \\ \hline $\Delta \lambda
_{ab}^{2}\left( T_{p_{c}}\right) /\lambda _{ab}^{2}\left(
T_{p_{c}}\right) $ & 0.11(5) & 0.15(6) & 0.15(5) \\ \hline
$^{16}T_{p_{c}}$(K) & 89.0(1) & 67.0(1) & 52.1(1) \\ \hline
$^{16}L_{c}$(A) & 9.7(4) & 14.2(7) & 19.5(8) \\ \hline
$^{18}T_{p_{c}}$(K) & 89.0(1) & 66.0(2) & 51.0(2) \\ \hline
$^{18}L_{c}$(A) & 10.9(4) & 16.0(7) & 22.6(9) \\ \hline
\end{tabular}
\end{center}

Table I: Finite size estimates for the relative changes of
$L_{c}$, $T_{p_{c}}$ and $\lambda _{ab}^{2}\left( T_{p_{c}}\right)
$ upon oxygen isotope exchange for
Y$_{1-x}$Pr$_{x}$Ba$_{2}$Cu$_{3}$O$_{7-\delta }$\cite {tsrkhk}.

\bigskip

To appreciate the implications of these estimates, we note again
that for fixed Pr concentration the lattice parameters remain
essentially unaffected \cite{conder,raffa}. Accordingly, an
electronic mechanism, without coupling to local lattice
distortions implies $\Delta L_{c}=0$. On the contrary, a
significant change of $L_{c}$ upon oxygen exchange requires local
lattice distortions involving the oxygen lattice degrees of
freedom and implies with Eq.(\ref{eq8}) a coupling between these
distortions and the superfluid. A glance to Table I shows that the
relative change of the grains along the $c$-axis upon oxygen
isotope exchange is significant, ranging from $12$ to $16\% $,
while the relative change of the inflection point at $T_{p_{c}}$,
or the transition temperature, is an order of magnitude smaller.
For this reason the significant relative change of $L_{c}$ at
fixed Pr concentration is accompanied by essentially the same
relative change of $\lambda _{ab}^{2}$, which probes the
superfluid. This uncovers unambiguously the existence and
relevance of the coupling between the superfluid and lattice
distortions, involving the oxygen lattice degrees of freedom.
Furthermore the substantial isotope effect on the in-plane
penetration depth at $T=T_{p_{c}}$ extends the evidence for the
failure of the Migdal-Eliashberg (ME) theory of the
electron-phonon interaction, predicting $1/\lambda ^{2}$ to be
independent of the ionic masses\cite{migdal}, to finite
temperature. Although the majority opinion on the mechanism of
superconductivity in the cuprates is that it occurs via a purely
electronic mechanism involving spin excitations, and lattice
degrees of freedom are supposed to be irrelevant, the relative
isotope shifts $\Delta L_{c}/L_{c}\approx \Delta \lambda
_{ab}^{2}/^{16}\lambda _{ab}^{2}\approx 14\%$ and $\Delta
d_{s}/d_{s}\approx 3\%$ uncover clearly the existence and
relevance of the coupling between the superfluid and local lattice
distortions. Recent site-selective oxygen isotope
($^{16}$O/$^{18}$O) effect measurements of the in-plane
penetration depth in
Y$_{0.6}$Pr$_{0.4}$Ba$_{2}$Cu$_{3}$O$_{7-\delta }$, using the
muon-spin rotation ($\mu $SR) technique show that the distortions
arise from the oxygen sites within the CuO$_{2}$ planes (100 \%
within error bar)\cite {rksite}. Potential candidates are then the
Cu-O bond-stretching-type modes showing a temperature dependence,
which parallels that of the superconductive order
parameter\cite{chung}.

To summarize we observed remarkable consistency between the
scaling properties of the experimental data for a variety of
cuprates and those characterizing 2D-QSI transitions. The
important implication there is that in cuprates a no vanishing
transition temperature and superfluid density in the ground state
are unalterably linked to a finite anisotropy. Furthermore, the
oxygen isotope effect on the in-plane penetration depth and the
spatial extent of the superconducting grains revealed the coupling
between local lattice distortions and superfluidity, while the
lattice parameters remain essentially unaffected. These findings
raise serious doubts that 2D models \cite{anderson}, neglecting
granularity and local lattice distortions are potential candidates
to explain superconductivity in cuprates.

\bigskip
\acknowledgments The author is grateful to D. Di Castro, R.
Khasanov, K.A. M\"{u}ller, and J. Roos for very useful comments
and suggestions on the subject matter. This work was partially
supported by the Swiss National Science Foundation and the NCCR
program \textit{Materials with Novel Electronic Properties}
(MaNEP) sponsered by the Swiss National Science Foundation.

\end{document}